\documentclass[aps,prc,showpacs,twocolumn]{revtex4}
\usepackage{graphicx}
\usepackage{dcolumn} 
\usepackage{bm}
\usepackage{longtable}
\usepackage{hyperref}
\usepackage{natbib}
\usepackage{amsmath}
\begin{document}
\title{Effect of first forbidden decays on the shape of neutrino spectra}
\author{Dong-Liang Fang$^{a,b}$ and B. Alex Brown$^{a,b,c}$}
\affiliation{$^a$National Superconducting Cyclotron Laboratory, Michigan State
University, East Lansing, Michigan 48824, USA}
\affiliation{$^b$Joint Institute for Nuclear and Astrophysics, Michigan State
University}
\affiliation{$^c$Department of Physics and Astronomy, Michigan State University,
East Lansing, MI 48824, USA}
\begin{abstract}
We examine the effect of First Forbidden (FF) decays on $\beta$-decay neutrino spectra by performing microscopic nuclear structure calculations. By analyzing the FF decay branches of even-even nuclei we conclude that FF decays may be responsible for part of the missing neutrinos in the so called "Reactor Neutrino Anomaly". Further calculations and more experimental data are needed for a firm conclusion.
\end{abstract}
\pacs{14.60.Lm,21.60.-n, 23.40.Bw}
\maketitle
\section{introduction}

The "reactor anti-neutrino anomaly" is the observation that the average of the experimentally determined reactor anti-neutrino flux at reactor-detector distances less than 100 m accounts for only 0.946 $\pm$ 0.023 of the theoretical expectation \cite{Abe11,MFL11,MLF11}. One of the explanations for this anomaly is that standard neutrinos that carry the weak nuclear charge can oscillate into a ``sterile" neutrino that does not contain a weak nuclear charge and escapes detection. This is one of the few types of experiments that could be sensitive to the sterile neutrinos. It is important to verify that the shape of the calculated anti-neutrino spectrum is correct.



In \cite{Hub11} the effects of various corrections to the expected neutrino spectra were examined, but it was found that these corrections could not explain the anomaly. However, in  \cite{Hub11} only allowed decays were analyzed and the First Forbidden(FF) decay was left out. In \cite{HFG13} the contribution of FF decays were examined by assuming some virtual FF branches in the decay, and it was found that the results could match the magnitude of the missing flux. However, the actual FF decays are much more complicated than the form assumed in \cite{HFG13}. In this work, we will exam the effect of realistic FF decay from microscopic nuclear structure calculations, and make an estimation of their effect on the neutrino spectra.   

For the neutron-induced fission accumulated yield distributions one finds two peaks
for the fission products. One of these is centered near  $^{94}$Sr and another is centered
near $^{140}$Xe. Both or these regions could have comparable amount of FF branches. For the region
of  $^{140}$Xe calculations with the Quasi-Particle Random Phase (QRPA) method 
were carried out \cite{FBS13}, and reasonable agreement between
experiment and theory 
for half-lives and $log$ft values were obtained. Also shell-model (SM) calculations for some nuclei
in the Xe region
can be carried out in a moderately large basis with previously derived Hamiltonian.
Thus, in this work we will focus on applying the QRPA and SM methods
in this Xe region to investigate the effect of FF beta decay on the shape of the
neutrino spectra.

This article is arranged as follows. First we give some background on $\beta$-decay, especially the FF decays, and the nuclear structure theories we used in our calculations. Then we present the calculated results with comparisons to experiments and the corresponding neutrino spectra. Conclusions are given at the end.
\section{Theory of $\beta$-decay }
The decay rate for $\beta$-decay can be written generally as \cite{SYK11,ZCC13}:
\begin{eqnarray}
\lambda = \ln2/t_{1/2}=\sum_{i} \lambda_{i}.
\end{eqnarray}
With the conventions and numerical constants used in \cite{SYK11,ZCC13} one obtains:
\begin{eqnarray}
f&=&8896\ s^{-1}\ \lambda \nonumber\\
&=& \sum_{i}\int_{1}^{\omega_{0i}} C(\omega) F(Z,\omega)p\omega(\omega_{0i}-\omega)^2 d\omega.
\end{eqnarray}
Here $\omega\equiv E_e/m_e$ is the energy of the emitted electron in the units of electron mass, $\omega_0$ is the $\beta$-decay energy in the unit of electron mass, and $p=\sqrt{\omega^2-1}$ is the momentum of the electron. $F(Z,\omega)$ is the Fermi factor which takes into account of 
nuclear charge on the shape of the spectra for the emitted electron.  

The nuclear structure dependence on the shape of the emitted leptons is contained in 
$C(\omega)$. It has different $\omega$-dependencies for different kinds of decays that lead to the different spectra for emitted electron and neutrino. For allowed-decay, $C(\omega)$ is independent of $\omega$. For the FF decay, the dependence can be written in the form \cite{SYK11}:
\begin{eqnarray}
C(\omega)=K_0+K_1\omega+K_{-1}/\omega+K_2\omega^2.
\end{eqnarray}
For FF decays one has three different types of transitions associated with the change of spins, $\Delta \, J^{\pi}=0^-,1^-,2^-$, they have different matrix elements and $\omega$ dependencies:
\begin{eqnarray}
C^{\Delta_J=0}(\omega)&=&K_0+K_{-1}/\omega \nonumber \\
C^{\Delta_J=1}(\omega)&=&K_0+K_1\omega+K_{-1}/\omega+K_2\omega^2\nonumber \\
C^{\Delta_J=2}(\omega)&=&K_0+K_1\omega+K_2\omega^2
\end{eqnarray} 
The detailed expressions for the $K$'s can be obtained from \cite{SYK11,ZCC13}. For 0$^-$, there are three matrix elements $M_0^s$, $M_0^{s}{'}$ and $M_0^T$, for 1$^-$ one has five matrix elements involving $u$, $u'$, $x$, $x'$ and $y$, and for 2$^-$ just one matrix element $z$ is involved. The expressions for these matrix elements are given in \cite{SYK11}. In \cite{HFG13} only the $M_0^s$, $u$, $x$ and $z$ terms
were used for FF branches. Our additional terms result in some differences between our results
and those of \cite{HFG13}.

To get the electron or neutrino spectra, we take derivatives over the respective energies:
\begin{eqnarray}
\frac{dN_e}{d\omega}&=&N\frac{d\lambda_e}{d\omega}=C(\omega)F(Z,\omega)p(\omega_0-\omega)^2 \nonumber\\
\frac{dN_\nu}{d\omega_\nu}&=&N\frac{d\lambda_\nu}{d\omega_\nu}\\
&=&C(\omega_0-\omega_\nu)F(Z,\omega_0-\omega_\nu) \omega_\nu^2 \sqrt{(\omega_0-\omega_\nu)^2-1} \nonumber
\end{eqnarray}
The spectra for FF decays are different from that of allowed GT, and 
their shape
depends on the decay modes ($J^\pi$). 
To obtain the spectra
we need to know some detailed structure information for the $\beta$-active nuclei.

For the nuclear structure calculations, the configuration interaction model or shell model (SM) provides an exact solution within a model space for a restricted set of valence orbitals. Realistic shell-model Hamiltonians can be derived from renormalized interactions based on the nucleon-nucleon interaction with some empirical single-particle energies and modifications to reproduce
experimental binding energies and excitation energies. However, as the number of valence nucleons increase, the dimensions of the configurations increase drastically making the
calculations impossible.
Starting with a closed shell of  $^{132}$Sn one add nucleons in the ``$jj56$" model space that consists of  the five
($1g_{7/2}$, $2d_{5/2}$, $2d_{3/2}$, $3s_{1/2}$, $1h_{11/2}$) orbitals for protons
and the six ($1h_{9/2}$, $2f_{7/2}$, $2f_{5/2}$, $3p_{3/2}$, $3p_{1/2}$, $1i_{13/2}$) orbitals for neutrons.  
We can consider up to four neutrons and four protons in this $jj56$ model space. The
SM can be applied to the decays of nuclei with both even and odd numbers of protons or neutrons.

To obtain results over a wider region of the nuclear chart, one needs to use various approximations. One of these is the Quasi-particle Random Phase Approximations (QRPA) which assumes the excited states of the nuclei are small harmonic oscillations beyond the Hartree-Fock-Boglyubov (HFB) or BCS ground states. Only two quasi-particle excitations are considered in this approximation. By changing one neutron to one proton or {\it vice versa}, we obtain the spectra for odd-odd nuclei, this is the so-called pn-QRPA method \cite{HS67} which is usually used for charge exchange reactions as well as $\beta$-decay. The QRPA method
can only be applied to even-even nuclei.

\section{Results and Discussion}

\begin{table*}
\centering
\caption{List of excitation energies and spin-parities of the final states and the corresponding log$ft$ values from the experiments, the shell-model (SM) and pn-QRPA(QRPA) calculations for different nuclei, we are choosing here only important low-lying FF branches. The measured half-lives from \cite{NNDC} are presented here. The excitation energies are in the unit of MeV. For QRPA calculations compared with \cite{FBS13}, we have minor changes on the quenching (explained in the text) to make it much closer to the experimental results in this region for a better comparison.}
\begin{tabular}{|c|cc|ccc|ccc|ccc|}
\hline
                   &       &     &   & Exp. \cite{NNDC}& & & ShM& & &QRPA & \\
                   &  $J^{\pi}_i$& t(s) & $J^{\pi}_f$ & $E_{ex}$ & log$ft$ &$J^{\pi}_f$ 
                   & $E_{ex}$& log$ft$ & $J^{\pi}_f$ & $E_{ex}$ & log$ft$\\
\hline
                   & $0^+$   &   17.63  & $(1^-)$           & 0           & $>$6.7 
                                                               & $1^-$             & 0            &  6.85
                                                               &  $0^-$            & 0            &  6.37        \\
$^{136}$Te &             &            & $(0^-,1,2^-)$  & 0.222     & 7.23     
                                                               & $2^-$             & 0.095     & 7.37
                                                               & $1^-$             & 0.171     & 6.95         \\
                   &            &            & $(0^-,1)$        & 0.334     & 6.27     
                                                               & $0^-$             &   0.133   &  6.41  
                                                               &  $2^-$            & 0.194     & 7.89   \\ 
                   &            &            & $(0^-,1)$        & 0.631     & 6.28     
                                                              &    $1^-$          & 0.426     & 6.26    
                                                               & $2^-$            & 0.541     & 6.99         \\
                   &            &            & $(0^-,1,2^-)$& 0.738     & 7.57     
                                                               &    $2^-$        & 0.507     &  6.71      
                                                               &  $1^-$          & 0.747     & 6.13     \\
\hline
                   &  $0^+$    &13.6     &  $1^-,0^-$   & 0.080    & 6.14  
	                                                       &     &    &   
                                                                &  $0^-$         &  0          & 6.15        \\
$^{140}$Xe &               &            & $(0,1^-)$     & 0.515    & 6.82  
  	                                                       &     &    &   
                                                                & $1^-$          &  0.127   & 6.77       \\
                   &               &             & $0^{(-)},1^{(-)}$ & 0.653& 5.98     
	                                                       &     &    & 
                                                                & $2^-$          &  0.365   &  7.01     \\
                   &               &             & $(1,2^-)$    & 0.800    &$\approx$7.1
	                                                       &     &    & 
                                                                & $1^-$          &  0.586   & 6.05  \\
                   &               &             & $1^{(-)}$     & 0.966    & 6.77      
	                                                       &     &    & 
                                                                & $1^-$          &  1.353   &  6.75      \\
\hline
\end{tabular}
\label{ffrt}
\end{table*}

For the SM calculations  we used the NuShellX@MSU code \cite{ShX}.
The Hamiltonian for the $jj56$ model space
is taken from \cite{jj56}. For this model space, the spin-orbit partner of $h$ and $i$ levels are not included. As a result, a larger than average quenching is needed for the calculated Gamow-Teller
matrix elements.  
The truncation to $jj56$ will also require renormalization of the various FF operators.
The goal is to reproduce the experimental $log$ft values so that 
we will have realistic results for the neutrino spectra.

For the QRPA calculations we use a Hamiltonian similar to that used in \cite{FBS13}. We start
with the realistic G-matrix for CD-Bonn interaction, and then  introduce two renormalization parameters, $g_{ph}$ and $g_{pp}$, for the particle-hole and particle-particle channels, respectively. The fitting strategy for them, as well as for the quenching of factors for both GT and FF decay is explained in \cite{FBS13}. In this work we generally follow the previous work 
where we used $g_{A(V)} = 0.5\, g_{A0(V0)}$ for all types of transitions,
but slightly change some of the parameters
to better reproduce the $log$ft values in the Xe region: $g_A(1,2^-)=0.4 \, g_{A0}$, 
and $g_V(0^-)=0.6 \, g_{V0}$. The same quenching values are used for the SM calculations. 


\begin{figure*}
\includegraphics[scale=0.65]{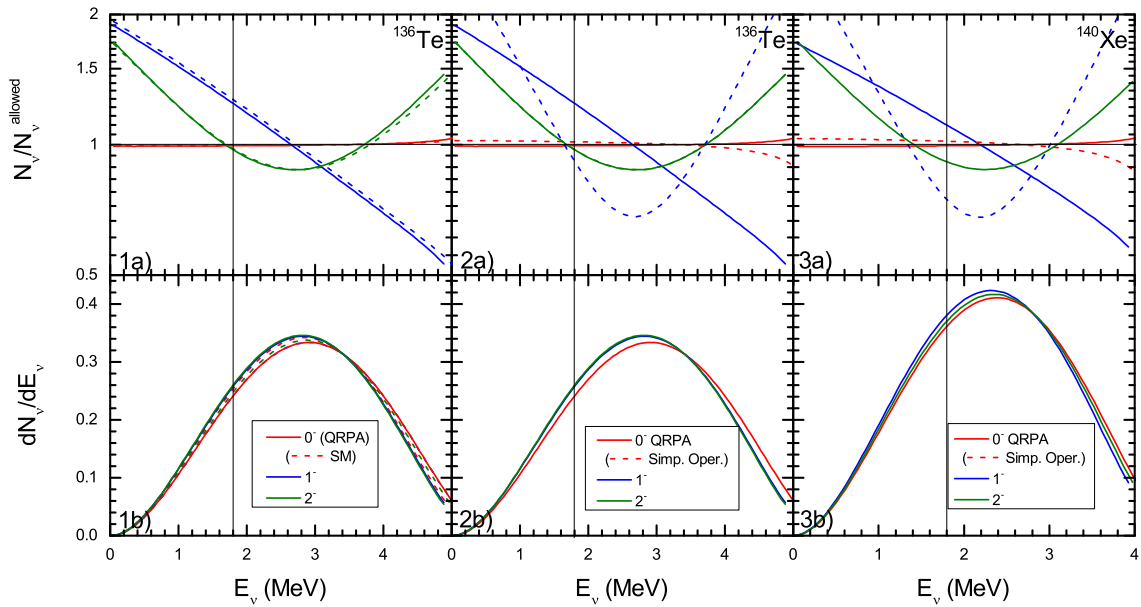}
\caption{(Color online) Neutrino spectra of low-lying FF decay branches for $^{136}$Te (left) and $^{140}$Xe (right) from SM (dashed lines in 1a) and 1b)) and QRPA (thick colors) calculations. The meanings of different line-styles are illustrated in the graph. For $^{140}$Xe, the Full FF means this calculations with matrix elements and phase space stated above in the text, meanwhile the ``Simp. FF" means the simplified matrix element used in \cite{HFG13}.}
\label{sgl}
\end{figure*}

In Table.\ref{ffrt} we present the comparisons of experimental results with the SM and QRPA methods for the two even-even nuclei. 
For $^{136}$Te, where the experimental data and both calculations are possible, we see good agreement among them. A one-to-one correspondence of most decay branches can be found between the SM calculations and the experimental results, the difference of the $log$ft values are within $0.2$ which means a factor of $1.5$ in the transition rates. The QRPA calculations agree with the shell model with differences for $log$ft values around $0.1-0.2$. 
Another even-even nucleus which has been measured is $^{140}$Xe. However, it is beyond the reach of our current SM computational capacity, so only QRPA results are shown. One finds that for this nucleus, the QRPA calculations are in good agreement with the measurement. 

\begin{table}
\centering
\caption{The percentage of the numbers of neutrinos which of the actual decay compared with the allowed shapes used in the simulation for single decay branches of $^{136}$Te and $^{140}$Xe, denoted by $\delta$ defined in text. The superscripts here are Q for QRPA and S for shell model, the subscript ``simp" means that we used the simplified FF matrix-elements used in \cite{HFG13}.}
\begin{tabular}{|c|ccccc|ccc|}
\hline
                   &  $E_{ex}^{Q}$& $\delta^{Q}$& $\delta_{simp.}^{Q}$ &  $E_{ex}^{S}$& $\delta^{S}$ &  $E_{ex}^{Q}$& $\delta^{Q}$ & $\delta_{simp.}^{Q}$\\
\hline
 $0^-$        &0.0       &  1.002  &0.995 & 0.133  & 1.001    & 0.0       & 1.003    &  0.990     \\
 $1^-_1$    &0.171   &  0.899  &0.929 &0.0      & 0.902    & 0.127   & 0.875    &  0.949     \\
 $1^-_2$    &0.747   &  0.938  &0.971 &0.426  & 0.933    & 0.586   & 0.919    &  0.981     \\ 
 $2^-_1$    &0.194   &  0.968  &          & 0.065  & 0.970    & 0.060   & 0.971    &                 \\
 $2^-_2$    &0.541   &  0.968  &         &0.507  & 0.982    & 0.365   & 0.976    &                 \\
\hline
\end{tabular}
\label{singsum}
\end{table}

As we have stated above, different decay channels may have different shapes due to different dependencies over energy $\omega$, so we need to investigate the effects of these decay channels on the neutrino spectra shape. For the odd-odd or odd-A nuclei there is usually mixing between different decay channels as $|J_i-J_f|\le \Delta J \le J_i+J_f$, but for even-even nuclei, because the ground states of the parent nuclei has always $J_i=0$, $\Delta J$ is unique for specific final state of daughter nuclei, there will be no mixing among different channels and it is easy to isolate different shape changes in different decay channels. 

In fig.\ref{sgl}, we compare the neutrino spectra shape changes relative to the allowed shape for different channels with different methods for two even-even nuclei ($^{136}$Te and $^{140}$Xe). For each nucleus we show the 0$^-$, 1$^-$ and 2$^-$ decay branches. The SM and QRPA methods agree well with each other.
For $0^-$ decays, the change of the spectra is small and it is a good approximation to treat the
$0^-$ decay as allowed decay.
For $1^-$ decay the change is large with the peak of the neutrino spectra shifted downwards.
This means that more neutrinos have less energy than expected from the previous simulation\cite{MFL11} using the allowed type of phase space.
For $2^-$ decay the behavior of the change to the shape is a bit different from that of $1^-$ as seen from fig.\ref{sgl} 
where the shape of the neutrino spectra for this decay branch is broadened.

We also make a comparison of the full microscopic calculations to the approximations made in \cite{HFG13}  
where 4 out of 9 matrix elements are used (affecting the 0$^-$ and 1$^-$ decays). 
For the 0$^-$ decay, the approximation used in \cite{HFG13} gives a result that is opposite to the full microscopic calculations, slightly shifting the neutrino spectra to lower energy.
For 1$^-$ decay, the approximation completely changes the behavior of the neutrino spectra. 
Due to the over simplified forms in \cite{HFG13}, the behavior of an overall shift of spectra to low energies disappears now. 
This comes from the fact that for simplified 1$^-$ decay in table I of \cite{HFG13} one of its matrix elements ($[\Sigma,r]^{1-}$ or $u$ in this work) has the same form as that for 2$^-$ decay ($[\Sigma,r]^{2-}$ or $z$ in this work).

There is similar behavior between $^{136}$Te and $^{140}$Xe. We would also expect the same behaviors of these FF decay channels in odd-mass or odd-odd nuclei since they have the same transition operators as the even-even nuclei. From the above results, we conclude that the inclusion of FF decays could eliminate 
the ``reactor anti-neutrino anomaly" if there are enough beta branches containing 1,2$^-$ transitions with suitable end-point energies, especially 1$^-$. However, if we examine the nuclear chart for the decay branching ratios, we find that 1,2$^-$ are usually accompanied with $0^-$ decays which usually have a much smaller $log$ft values (a stronger transition probability). This would reduce the overall changes to the spectra.

To quantify the change in the neutrino spectrum due to the change of phase space, we integrate over the spectra with the two phase spaces as follows, 
\begin{eqnarray}
\delta&=&\frac{1-n_{FF}(E< E_{t})}{1-n_{GT}(E< E_{t})} \nonumber\\
n_{I}(E< E_{t})&=&\int_{0}^{E_{t}}\frac{dN}{d E_{\nu}}(E_\nu)dE_{\nu}
\label{dele}
\end{eqnarray} 
 with $\int_{0}^{E_{end}}dN/d E_{\nu}(E_\nu)dE_{\nu}=1$.
$E_{t}$ is the energy needed to trigger the interaction $\bar{\nu}_e+p \rightarrow e^-+n$, and $E_{end}$ 
is the maximum energy of emitted neutrinos. 
The reduction in the number of low-energy neutrinos is given by $\Delta=1-\delta$.
The change depends on the end point energy $E_{end}$, which can be expressed as $Q_\beta-m_e-E_{ex}$. So we need precise excitation energies for the determination of neutrino spectra.
This result can then be compared with the value of the reactor neutrino anomaly to see if the lack of  
FF phase space factor in the simulation can explain the missing neutrinos. The results for single decay branches are listed in Table\ref{singsum}. 
A comparison between QRPA and shell model shows similarities for the ratio $\delta$, this agrees with Fig.\ref{sgl}. For the detailed values; the change $\Delta$ of the $0^-$ decay is negligible, for 1$^-$, $\Delta$ goes up to ten percent, and for 2$^-$, $\Delta$ is only 2-3 percent.

\begin{figure}
\includegraphics[scale=0.35]{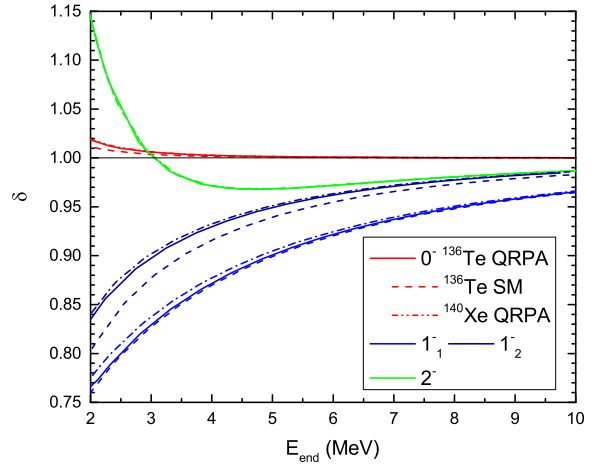}
\caption{(Color online) Dependence of the changes for the percentage of neutrino number $\delta$ (defined in text) on the end-point energies for several FF decay branches of $^{136}$Te(bold lines for QRPA calculations and dashed lines for SM calculations) and $^{140}$Xe(dashed-dot lines for QRPA calculation). Here we varying the Q values of the two nuclei above to see how the changes are related to the end-point energies of the decay branches.}
\label{dn}
\end{figure}

To obtain quantitative results on the dependence of the detailed changes on the end point energies of the decay branches, we vary the Q values in the calculations for the two nuclei $^{136}$Te and $^{140}$Xe. 
The results are plotted in fig.\ref{dn} where
one observes that to a large extent this relation is nucleus independent. 
$\Delta$ for the $0^-$ branches are near zero except below end-point energies of 3 MeV.
For small end-point energies, $\Delta$ is large due to the shape changes at the spectra tail; but these are not important since contributions of these branches to the total spectra are small, see Fig.3 of \cite{HFG13}. 
For $2^-$ decay the dependence of $\delta$ on the end-point energies are independent of $log$ft values since it has only one component. For end-point energies from 4-6 MeV, $\Delta$ is around $3-4\%$. 

However, for $1^-$ decays $\delta$ depends on both $E_{end}$ and $log$ft. To see this we also plot the $1^-_2$ decay branches for the two nuclei Fig.\ref{dn}.  Compared to $1^-_1$
the FF decays to the $1^-_2$ states
have smaller $log$ft values (Table.\ref{ffrt}) (i.e. they are stonger) 
and have smaller $\Delta$ values (Fig.\ref{dn}). 
The reason of this comes from the fact that the transition rates of $1^-$ are determined by five different components. They are combined to give the final decay rates, and their different combinations have different energy dependencies. 
At $E_{end}\sim 4-6$ MeV, $\Delta$ is $5-15\%$.
It was estimated in \cite{HFG13} that $30\%$ of the decay branches of the fission products are FF.
Thus, in the most extreme case where the FF is dominated by $\Delta J^\pi = 1^-$ 
the change of the neutrino spectrum could be as large as $\Delta = 4.5\%$.

\section{conclusion}
In this work, explicit analysis of $\beta$-decay neutrino spectra with inclusion of the 
first forbidden part has been performed. One finds that use of 
the allowed decay phase space factor results in a correction of up to about $\Delta=4.5\%$ due 
to $\Delta J^\pi=1^-$ FF transitions. 
An average over all types of FF transitions,
end-point energies and $log$ft values would result in a smaller value of $\Delta=1-2\%$. 
The finite size effects and  the weak magnetism corrections obtained in \cite{Hub11}
for the allowed (GT) decays are estimated to be $\Delta=2-3\%$.  
If the average branching ratios for all types of FF is estimated, they can be combined
with our results to obtain an improved correction for the shape of the
neutrino spectra. 

\begin{acknowledgements}
We would like to thank Prof. A. Hayes for useful discussions and helpful data.
This work was supported by the US NSF grants PHY-0822648 and PHY-1404442.
\end{acknowledgements}

\end{document}